\documentclass[prd,twocolumn,floatfix,preprintnumbers,letterpaper]{revtex4}
\usepackage{amssymb,amsbsy,bm,amsmath,psfrag,dsfont}
\usepackage{hyperref}
\usepackage{graphicx}
\usepackage{epsfig}

\begin{document}

\newcommand{\ba}{\begin{array}}
\newcommand{\ea}{\end{array}}
\newcommand{\nn}{\nonumber}
\newcommand{\no}{\noindent}
\newcommand{\la}{\lambda}
\newcommand{\si}{\sigma}
\newcommand{\vp}{\mathbf{p}}
\newcommand{\vk}{\vec{k}}
\newcommand{\vx}{\vec{x}}
\newcommand{\om}{\omega}
\newcommand{\Om}{\Omega}
\newcommand{\ga}{\gamma}
\newcommand{\Ga}{\Gamma}
\newcommand{\gaa}{\Gamma_a}
\newcommand{\al}{\alpha}
\newcommand{\ep}{\epsilon}
\newcommand{\app}{\approx}
\newcommand{\cO}{{\cal O}}
\newcommand{\A}{{\text A}}
\newcommand{\B}{{\text B}}
\newcommand{\AB}{{\text{AB}}}
\newcommand{\nc}{\newcommand}
\nc{\beq}{\begin{equation}}
\nc{\eeq}{\end{equation}}
\nc{\beqa}{\begin{eqnarray}}
\nc{\eeqa}{\end{eqnarray}}
\def\DS {D\!\!\!\!/}
\def\A { A_\mu (x) }
\def\DM {\DS_{\, (\mu)}}
\def\O {{\cal O}}
\def\ts{\thinspace}
\def\fpi{F}
\def\gsim{\mathrel{\rlap{\lower4pt\hbox{\hskip1pt$\sim$}}
    \raise1pt\hbox{$>$}}}       

\title{Does the BICEP2 Observation of Cosmological Tensor Modes Imply
an Era of Nearly Planckian Energy Densities?}
\author{Chiu Man Ho} \email{cmho@msu.edu}
\affiliation{Department of
Physics and Astronomy, Michigan State University, East Lansing, MI 48824, USA}
\author{Stephen~D.~H.~Hsu} \email{hsu@msu.edu}
\affiliation{Department of
Physics and Astronomy, Michigan State University, East Lansing, MI 48824, USA}
\date{\today}

\begin{abstract}

BICEP2 observations, interpreted most simply, suggest an era
of inflation with energy densities of order ($10^{16}\, {\rm GeV})^4$,
not far below the Planck density. However, models of TeV gravity with
large extra dimensions might allow a very different interpretation involving
much more modest energy scales. We discuss the viability of inflation
in such models, and conclude that existing scenarios do not provide
attractive alternatives to single field inflation in four dimensions.
Because the detection of tensor modes strengthens our confidence that
inflation occurred, it disfavors models of large extra dimensions, at least for the moment.

\end{abstract}
\maketitle

\section{Introduction}

Can the polarization of cosmic microwave background (CMB) photons shed
light on physics at the very highest energy scales? Obviously not
without the use of a theoretical framework, relying on specific prior
assumptions \cite{Amjad}. If one considers single field models of inflation in four
dimensions, BICEP2 suggests an era of very high energy density, not
far from the Planck density $\equiv (10^{19} \,{\rm GeV})^4$. We probe the robustness of this
conclusion by considering inflation scenarios in models of large extra
dimensions, where the maximum energy scale is only of order TeV.
In these models the effective strength of gravity depends on geometrical factors, and may have varied over cosmological time (i.e., if the size of extra dimensions was not constant).
Are there attractive alternative scenarios which do not require an era of nearly
Planckian energy density?

Single field inflation is perhaps the most economical and compelling framework for understanding the CMB
anisotropy \cite{Guth,Linde}. While there are many models for single field inflation, most of them
predict nearly scale-invariant scalar and tensor primordial fluctuations. The power spectra for scalar and tensor
perturbations are respectively \cite{Weinberg}
\beqa
\label{scalar}
P_S &\approx& \frac{1}{8\,\pi^2}\,\left(\frac{H_\ast^2}{\epsilon\, M_P^2}\right)\;, \\
\label{tensor}
P_T &\approx& \frac{2}{\pi^2}\,\left(\frac{H_\ast^2}{M_P^2}\right) \;,
\eeqa
where $H_\ast$ is the Hubble rate during inflation, $M_P = 1/\sqrt{8\,\pi\, G} \approx 2.44 \times 10^{18}\, \textrm{GeV}$ is the reduced
Planck scale and $\epsilon$ is the standard slow-roll parameter:
\beqa
\label{slowroll}
\epsilon = \frac{1}{2 \, M_P^2} \frac{\dot{\phi}^2}{H_\ast^2} \;.
\eeqa
According to the measurement by the PLANCK satellite \cite{PlanckInflation}, we have
$P_{S} \approx 2.19 \times 10^{-9}$ and this implies that
\beqa
\label{Hstar}
H_\ast = 1.13 \times 10^{14} \,\textrm{GeV} \, \left(\,\frac{r}{0.2}\,\right)^{1/2}\;,
\eeqa
where $r = P_T / P_S$ is the tensor-to-scalar ratio. $H_\ast$ is related to the energy density at the time of inflation
$V$ through
\beqa
\label{Friedmann}
H_\ast^2 = V / 3 M_P^2\;,
\eeqa
and therefore
\beqa
\label{EnergyDensity}
V^{1/4} = 2.19 \times 10^{16} \,\textrm{GeV} \, \left(\,\frac{r}{0.2}\,\right)^{1/4}\;.
\eeqa

The recent BICEP2 measurement of the tensor modes from large angle CMB B-mode polarization indicates
a tensor-to-scalar ratio \cite{BICEP2}
\beqa
\label{ratio}
r = 0.2^{+0.07}_{-0.05}\;.
\eeqa
This result, if confirmed, suggests that $H_\ast \sim 10^{14}\, \textrm{GeV}$, implying
an era of inflation with energy density $V \sim (10^{16}\, \textrm{GeV})^4$.

Using $r = P_T / P_S$, Eq. \eqref{slowroll} and the definition of the number of e-foldings as $dN = H_\ast \,dt$,
we obtain
\beqa
\frac{1}{M_P} \,\frac{d\phi}{d N} = \sqrt{2\, r}\;.
\eeqa
For the relevant CMB anisotropy multipoles within the Hubble radius, the
corresponding number of e-foldings is $\Delta N \approx 4.6$. Thus,
\beqa
\frac{\Delta \phi}{M_P} \gtrsim \left(\,\frac{r}{0.02}\,\right)^{1/2}\;,
\eeqa
assuming that the total number of e-foldings is $N \gtrsim 50$. This means that the observation of primordial
gravitational waves as large as $r \sim 0.2$ requires Planckian excursions of the inflaton field during the slow-roll phase. Flatness of the potential for such large excursions implies strong constraints on higher dimension operators involving the inflaton \cite{Lyth}. We refer the reader to discussions in the literature \cite{EFT}, especially the cautionary remarks made in \cite{Riotto}.

Note that these considerations also affect models of TeV gravity: again in
these models $r$ cannot be large unless the energy density during
inflation is close to the scale of quantum gravity. Flatness of the
inflaton potential is then sensitive to higher dimension operators
induced by quantum gravity.

\section{Large Extra Dimensions}

The possibility that large extra dimensions might explain the large hierarchy between the Planck and electroweak scales has been extensively investigated. The most popular models in this vein are Arkani-Hamed-Dimopoulos-Dvali (ADD) \cite{ADD},
Randall-Sundrum I (RS1) \cite{RS1} (note Randall-Sundrum II (RS2) \cite{RS2} does not solve the hierarchy problem and we do not focus on it here).  In these models the Standard Model (SM) particles are confined to a 3-brane, while the gravitons can also
propagate in the compact extra dimensions. The metric describing this kind of $D$-dimensional spacetime can be written as
\beqa
\label{metric}
ds^2 = e^{A(y)}\,dx_4^2 - g_{\alpha \beta} \,dx^\alpha \, dx^\beta \;,
\eeqa
where $dx_4^2 = \eta_{\mu\nu}\,dx^\mu\,dx^\nu $ is the standard 4-dimensional Minkowski line element and the compact
extra dimensions are parameterized by the coordinates $y$ with the metric elements $g_{\alpha \beta}$. The function $e^{A(y)}$
is called the warp-factor. For $A(y) \neq 0$, we have warped extra dimensions (RS1). For $A(y) \rightarrow 0$, the metric in Eq. \eqref{metric}
reduces to the case with flat extra dimensions (ADD).

The 4-dimensional Newton constant $G_4 = 1/ 8 \,\pi\, M_P^2 $ is related to the $D$-dimensional one $G_D$ by
\beqa
G_4^{-1} = G_D^{-1}\, V_{D-4}\;,
\eeqa
where $G_D^{-1}$ characterizes the Planck scale in $D$-dimension and
$V_{D-4}$ is the volume of the extra dimensions given by
\beqa
\label{volume}
V_{D-4} =  \int\, d^{D-4} \, \sqrt{g}\,\, e^{A(y)}\;.
\eeqa

\section{Inflation Models}

In both ADD and RS1 the fundamental Planck scale $M_\ast$ is of order TeV.
For RS1, we have TeV and Planck branes, with SM particles on the TeV brane.
A quick inspection of Eq. \eqref{scalar}, Eq. \eqref{tensor} and Eq. \eqref{Friedmann} suggests that one might simply replace $M_P$ by $M_\ast$ so that
$V$ need not be as large as ($10^{16}\, {\rm GeV})^4$. That is, a much lower energy density might be able to produce the large value of $r$ observed by BICEP2. However, a more detailed analysis of inflation in these models is necessary. We divide the scenarios into two categories, depending on whether the inflaton is confined to the (TeV) brane.

\subsection{Brane Inflaton}

Regardless of whether the large extra dimensions are flat or warped, for an inflaton confined to the (TeV) brane, the relevant Newton's constant would be $G_4$. This means that the power spectrum of the tensor fluctuations
would follow Eq. \eqref{tensor} and we would need an energy density of order ($10^{16}\, {\rm GeV})^4$ to generate the
observed tensor modes by BICEP2. However, this energy scale is much larger than the TeV scale of quantum gravity in both of ADD and RS1. Hence, at best, quantum gravitational effects would come into play in studying such high energy densities and the behavior of the model is incalculable. Most likely, it does not result in the desired inflationary behavior.

The reasoning in the previous paragraph may be evaded in a well-known ADD scenario \cite{ADDInflation} where the extra dimensions are assumed to be static with size $1/ \textrm{TeV} \sim 2\times 10^{-19} \,\textrm{m}$ during inflation; these extra dimensions only become large post-inflation. In this case, one can replace $M_P$ by $M_\ast$, although the specific relations for the scalar and tensor perturbations are more complicated.
In any case, this model has a fine-tuning problem: unless inflation ends with the scale factor of the extra dimension(s) (i.e., the radion) almost precisely at the minimum of its effective potential, residual oscillations will over close the universe. As explained in \cite{Radions}, this problem cannot be solved by another epoch of limited inflation, as originally proposed in \cite{ADDInflation}.


Taking the inflaton on the (negative tension) TeV brane of RS1 is known to be problematic as the slow-roll
inflation destabilizes the brane separation \cite{MazumdarRSbrane} -- the radion associated with the relative separation
of the TeV and Planck branes. Therefore, unlike \cite{ADDInflation}, one cannot assume that the extra dimension is static
with size $1/ \textrm{TeV} \sim 2\times 10^{-19} \,\textrm{m}$ during inflation and thereby
replace $M_P$ by $M_\ast$ in the equations describing the expansion. In this case, the BICEP2 observation would require an energy scale much larger than the scale of quantum gravity in RS1.



\subsection{Bulk Inflaton}

Next we turn to models in which the inflaton lives in the bulk. We summarize a number of models below, each of which has its own difficulties.

The scenario in \cite{Mohapatra} considers an ADD bulk inflaton. It is assumed that the
extra dimensions are already stabilized (by an unknown dynamical mechanism) at their large size during inflation. However, the
purpose of inflation is to eliminate the need for special cosmological initial conditions and thus this model is not, in our opinion,
a successful inflation model.



The scenario in \cite{MazumdarADDbulk} allows both the 4-dimensional brane spacetime and the flat extra dimensions to
inflate simultaneously. However, it has been pointed out by
\cite{ADDInflation} that the stabilized size of the extra dimensions in \cite{MazumdarADDbulk} is unacceptably large and
a scale-invariant spectrum of density perturbations is impossible (see footnote 1 in \cite{ADDInflation}).

Perhaps the most economical scenario within the ADD framework is to let the radion be the inflaton. However,
for radions with simple polynomial potentials, the potential
must be of order $(10^{-4} \,M_P)^4 \sim (10^{14}\,\textrm{GeV})^4 \gg M_\ast^4 \sim (\textrm{TeV})^4$ in order to obtain the observed density perturbations \cite{ClineADD}.

A bulk inflaton with a quadratic potential in RS1 has been investigated in \cite{ClineRS1}. The magnitude of density perturbations
turns out to be suppressed by $\textrm{TeV}/M_P$ and is therefore not consistent with observations. Thus, the simplest chaotic
inflation in RS1 is ruled out if the inflaton is a bulk field.

Finally, we note that the RS1 scenario requires a radion stabilization mechanism to avoid a variety of cosmological problems \cite{Csaki}. Once a radion stabilization mechanism (such as the Goldberger-Wise mechanism \cite{Wise}) is introduced, the conventional cosmology is recovered at low temperatures if the radion is stabilized at the weak scale \cite{CsakiRandall}. The radion would have been an economical candidate of inflaton, but the Goldberger-Wise mechanism does not give it a flat potential. We do not know of any attractive inflation scenario incorporating radion stabilization that predicts a large tensor mode.

\section{Conclusions}

CMB tensor modes cannot be said to probe ultrahigh energies such as $10^{16}$ GeV without
additional theoretical assumptions. However, they are plausibly sensitive to energies not far below the scale of quantum gravity. There is significant uncertainty in the value of this scale, as illustrated in our discussion of models with large extra dimensions. The key question that arises is whether inflationary scenarios exist in these models which are as attractive as single field inflation in four dimensions. In our view, existing inflation scenarios in models with large extra dimensions are less
appealing than single field scenarios in four dimensions. In fact, they seem to be far more problematic.

If the BICEP2 tensor mode results are confirmed by
experiments such as PLANCK, confidence in inflationary cosmology
will increase significantly.  Therefore, in the absence of improvements in large extra-dimensional scenarios, confirmation of
BICEP2 will disfavor large extra dimensions and suggest very high energy densities in the early universe.

On a positive note, if attractive TeV gravity inflationary scenarios are found (consistent with large tensor modes), they may provide a means to restrict the zoo of large extra dimension models. This category of models has long suffered from the diverse set of possibilities allowed by arbitrary assumptions concerning dynamics in the extra dimensions.

\section{Acknowledgements}

This work was supported by the Office of the Vice-President for Research and Graduate
Studies at Michigan State University.



\end{document}